\documentclass[12pt]{article}
\usepackage[dvips]{graphicx}
\def\doublespace{\lineskip      .25 ex\baselineskip 3.0
ex\lineskiplimit 0 ex\parskip 1.0 ex plus.50 ex minus .25 ex}%

\begin{document}
\doublespace

\title{A detailed discussion of superfield \\
supergravity prepotential
perturbations \\ in the superspace of the \\
$AdS_5/CFT_4$ correspondence}

\author{ J.
Ovalle\footnote{jovalle@usb.ve}
 \\
\vspace*{.25cm}\\
 Departamento de F\'{\i}sica, \\ Universidad Sim\'on Bol\'{\i}var, \\ Caracas,
 Venezuela.
 }
\date{}
\maketitle
\begin{abstract}

This paper presents a detailed discussion of the issue of
supergra\-vity perturbations around the flat five dimensional
superspace required for manifest superspace formulations of the
supergravity side of the AdS${}_{5}$/CFT${}_4$ Correspondence.

\end{abstract}

pacs 04.65.+e
\newpage
\section{Introduction}

The importance of deeply understanding the superspace
geometry of five dimensions, has received attention during the
last few years \cite{JimLubna}, \cite{kuzenko1}, \cite{kuzenko2},
motivated mainly by the postulate of the AdS/CFT correspondence
\cite{Maldacena}.  In this respect, the study of supergravity
theories represents an unavoidable issue \cite{kuzenko3},
\cite{kuzenko4}, \cite{kuzenko5}, even more keeping in mind
the existence of the supergravity side of the $AdS_5/CFT_4$
correspondence.  Indeed the works of \cite{kuzenko4,kuzenko5}
present complete nonlinear descriptions of such superspaces based
on a particular choice of compensators.  It has long been known
\cite{SFSG}, that the superspace geometry changes when different
compensators are introduced.  So one of the goals of the current
study is to begin the process of looking at what features of the
work of  \cite{kuzenko4,kuzenko5} are universal (i.e. independent
of compensator choice).

When the superspace approach is used, the conventional
representation for Grassmann variables for $SUSY$ $D=5$, ${\cal N}=1$
(often denominated ${\cal N}=2$) considers these variables obeying a
pseudo-Majorana reality condition, then the spinor coordinates are
dotted with an $SU(2)$ index.  Thus the conventional approach first
doubles the number of fermionic coordinates, by the introduction of
the $SU(2)$ index, then halves this number by the imposition of the
pseudo-Majorana reality condition.  However, as noted previously
\cite{JimLubna}, there is no fundamental principle that demands
the use of Majorana-symplectic spinors for describing the
fermionic coordinated.  Indeed, it was demonstrated in
\cite{JimLubna} that complex 4-components spinors provide an
adequate basis for describing such a fermionic coordinate.  Building
on this previous work, in this paper it will be shown that it is possible to develop
successfully a geometrical approach to five dimensional ${\cal N}=1$
supergravity theory, using this unconventional representation for
Grassmann variables.

~~ The geometrical approach to supergravity involves calculating
fields strengths to determine the form of the torsions and
curvatures of the theory. With this information in hand, we can
set constraints such that the super spin connections and the
vector-supervector component of the inverse supervierbein become
dependent variables of the theory. Once this is accomplished, the
route to deducing the prepotentials of the supergravity theory
are opened. On the other hand, there is an
alternative approach which is based in the torsion superfield
$C_{AB}^{\;\;\;\;\;C}$ associated to the superspace derivative
$E_A$, namely, the super-anholonomy
$[E_A,E_B\}=C_{AB}^{\;\;\;\;\;C}E_C.$ Using the superspace
derivative $\nabla_A$ to calculate the super (anti)-commutator
$[\nabla_A,\nabla_B\}$, we will be able to write all
super-torsion components in terms of the anholonomy and the
spinorial connection. Through a  choice of suitable constraints on
superspace through some super-torsion components, we will be able
to write the spin connection superfield in terms of the
anholonomy, eliminating by this way the spin connection as
independent fields. Once this is accomplished, the linearized
theory is considered through perturbations around the flat
superspace. In this linearized regime, all super holonomy
components can be obtained in terms of semi-prepotentials.
Hence the torsion and curvature of the theory are determinated in
terms of these semi-prepotentials.


\section{Superspace geometry: the unconventional \\ representation }

$~~~$
Let us start considering the supercoordinate
$Z^{A}=(x^m,\theta^{\mu},\bar{\theta}^{\mu})$, where the bosonic
and fermionic coordinates are given respectively by $x$ and
$\theta$, where $m=0,...,4$ and $\mu=1,...,4$. As already
mentioned, unlike the conventional representation for Grassmann
variables for $SUSY$ $D=5$, ${\cal N}=1$, the unconventional
representation for the Grassmann variables
$(\theta^{\mu},\bar{\theta}^{\mu})$ given in \cite{JimLubna}
where there is not an $SU(2)$ index appended to the spinor coordinates of
the superspace, will be used.

Under this unconventional spinorial representation for $SUSY$
$D=5$, ${\cal N}=1$, the spinorial supercovariant derivatives is given by
(for all details concerning this algebra, see \cite{JimLubna})
\begin{eqnarray}
\label{lub1}
D_\mu=\partial_\mu+\frac{1}{2}(\gamma^{m})_{\mu\nu}C_\sigma^{\;\;\nu}\bar{\theta}^{\sigma}
\partial_m,\hspace{1cm}\bar{D}_\mu=\bar{\partial}_\mu-\frac{1}{2}(\gamma^m)_{\mu\sigma}\theta^\sigma\partial_m,
\end{eqnarray}
which satisfies the algebra
\begin{eqnarray}
\label{lub2}
\{D_{\mu},\bar{D}_{\nu}\}=({\gamma}^m)_{{\mu}{\nu}}{\partial}_m,\hspace{0.5cm}\{D_{\mu},D_{\nu}\}=\{\bar{D}_{\mu},\bar{D}_{\nu}\}=0.
\end{eqnarray}
In order to construct the $SUGRA$ $D=5$, ${\cal N}=1$ version associated
with this representation, it is necessary to consider the
supervector derivative $\nabla_A$, a superspace supergravity
covariant derivative, which is covariant under gene\-ral
supercoordinate and superlocal Lorentz groups, given by
\begin{eqnarray}
\label{lub3}
\nabla_A=E_A+\Upsilon_A;\;\;\;\Upsilon_A\equiv\frac{1}{2}\omega_{Ac}^{\;\;\;\;d}M_d^{\;\;c}+\Gamma_AZ,\hspace{0.5cm}A=(a,{\alpha},\bar{{\alpha}})
\end{eqnarray}
This supergravity superderivative is written in terms of a
superderivative $E_A$, a spin super connection
$\omega_{Ac}^{\;\;\;\;d}$, the Lorentz generator $M_d^{\;\;c}$, a
central charge super connection $\Gamma_A$ and the central charge
generator $Z$. It can be seen there is an absence of any $SU(2)$
connection, which is characteristic of the unconventional
fermionic representation considered here.

We should also mention one other possibility (though we will not study
it in this work).  Since the bosonic dimension is five, it follows that the
field strength of a 5-vector gauge field is thus a two-form.  Hence in superspace
there must be a super two-form field strength (as appears in (2.5) below).
However, by Hodge duality, there should be expected to be a formulation of supergravity here where the central charge connection $\Gamma_A$
can be set to zero and instead there is introduced a super two-form
$\Gamma_{A B}$.

The superspace derivative $E_A$ is given through the super
vielbein $E_A^{\;\;M}$ by
\begin{equation}
\label{lub4}E_A=E_A^{\;\;M}D_M,
\end{equation}
with $D_M$ being the supervector (this is the flat situation)
which components are $({\partial}_m,D_\mu,\bar{D}_\mu)$ satisfying
the algebra (\ref{lub2}). The supertorsion $T_{AB}^{\;\;\;\;\;C}$,
curvature superfield $R_{AB{c}}^{\;\;\;\;\;\;\;d}$ and central
charge superfield strength are given through the algebra
\begin{equation}
\label{lub5}
[\nabla_A,\nabla_B\}=T_{AB}^{\;\;\;\;\;C}\nabla_C+\frac{1}{2}R_{AB{c}}^{\;\;\;\;\;\;\;d}M_d^{\;\;\;c}+F_{AB}Z.
\end{equation}
Again, there is no curvature associated with  $SU(2)$ generators.

The anholonomy superfield $C_{AB}^{\;\;\;\;\;C}$ associated to the
superspace derivative $E_A$ is given by
\begin{equation}
\label{anholonomy}
[E_A,E_B\}=C_{AB}^{\;\;\;\;\;C}E_C.
\end{equation}
This superfield structure will play a fundamental role in our analysis.

The first step will be to calculate the super (anti)-commutator
(\ref{lub5}) using the superspace derivative (\ref{lub3}). By this
way we will be able to write all super-torsion components in terms
of the anholonomy and the spinorial connection. Then choosing
suitable constraints on superspace through some super-torsion
components, we will be able to write the spin connection
superfield in terms of the anholonomy, eliminating thus the spin
connection as independent fields. Finally eliminated the spin
connection, the next step will be to obtain a specific form for
all components of the anholonomy superfield. To carry out this, it
will be necessary to provide an specific structure to the super
vielbein $E_A^{\;\;M}$. This structure will be based in all ``fundamental" geometric objects which appear in $D=5\;{\cal N}=1$ SUSY.
These fundamental objects are the following spinor metric and
gamma matrices
\begin{equation}
\label{fundamental}
\eta_{{\alpha}\beta};\hspace{1cm}({\gamma}^a)_{\alpha}^{\;\;\beta};\hspace{1cm}(\sigma^{ab})_{\alpha}^{\;\;\beta}.
\end{equation}
This will be explained in detail later.

 Let's start now with the first step, which is to calculate the super (anti)-commutator
(\ref{lub5}) using the superspace derivative (\ref{lub3}) and then
identify all super-torsion components. The super (anti)-commutator
can be written as
\begin{eqnarray}
\label{tor2}
[\nabla_A,\nabla_B\}=[E_A,E_B\}+[E_A,\Upsilon_B\}+[\Upsilon_A,E_B\}+[\Upsilon_A,\Upsilon_B\},
\end{eqnarray}
using the anholonomy definition we have
\begin{equation}
\label{tor3}
[\nabla_A,\nabla_B\}=C_{AB}^{\;\;\;\;\;C}E_C+[E_A,\Upsilon_B\}+[\Upsilon_A,E_B\}+[\Upsilon_A,\Upsilon_B\}.
\end{equation}
Hence finally we have the explicit form of the algebra, given by
(see appendix to details)
\begin{eqnarray}
\label{ALG1}
[\nabla_a,{\nabla}_b]&=&C_{a{b}}^{\;\;\;C}\nabla_C-{\omega}_{[a{b]}}^{\;\;\;\;\;c}\nabla_c
\nonumber
\\ \nonumber
&&+\frac{1}{2}\left[-C_{a{b}}^{\;\;\;C}\omega_{C}^{\;\;\;ef}+E_{a}{\omega}_{{b}}^{\;\;ef}
-{E}_{b}\omega_{{a}}^{\;\;ef}-
\omega_{{[a{}}\;\;}^{\;\;\;ec}\;{\omega}_{{b]c}}^{\;\;\;\;f}\right]M_{fe}
\\ \nonumber
&&+\left[-C_{a{b}}^{\;\;\;\;C}{\Gamma}_C+E_{a}{{\Gamma}}_{b}-{E}_{b}{\Gamma}_{a}\right]Z.\\
\end{eqnarray}
\begin{eqnarray}
\label{ALG2}
\{\nabla_{\alpha},\nabla_{\beta}\}&=&C_{{\alpha}{\beta}}^{\;\;\;\;C}\nabla_C+\frac{\imath}{4}\left[\omega_{{\alpha}}^{\;\;cd}({\sigma}_{dc})_{{\beta}}^{\;\;{\gamma}}
+\omega_{{\beta}}^{\;\;cd}({\sigma}_{dc})_{{\alpha}}^{\;\;{\gamma}}\right]\nabla_{\gamma}\nonumber
\\ \nonumber
&&+\left[-\frac{1}{2}C_{{\alpha}{\beta}}^{\;\;\;\;\;C}\omega_{Cc}^{\;\;\;\;\;d}+\frac{1}{2}E_{{\alpha}}\omega_{{{\beta}}c}^{\;\;\;\;d}
+\frac{1}{2}E_{{\beta}}\omega_{{{\alpha}}c}^{\;\;\;\;d}+
\omega_{{{\alpha}}\;\;c}^{\;\;\;b}\;\omega_{{{\beta}}b}^{\;\;\;\;d}\right]M_{d}^{\;\;c}
\\ \nonumber
&&+\left[-C_{{\alpha}{\beta}}^{\;\;\;\;\;C}{\Gamma}_C+E_{{\alpha}}{\Gamma}_{{\beta}}+E_{{\beta}}{\Gamma}_{{\alpha}}\right]Z,\\
\end{eqnarray}
\begin{eqnarray}
\label{ALG3}
\{\nabla_{\alpha},\bar{\nabla}_{\beta}\}&=&C_{{\alpha}\bar{{\beta}}}^{\;\;\;\;\;C}\nabla_C+\frac{\imath}{4}\bar{\omega}_{{\beta}}^{\;\;cd}({\sigma}_{dc})_{{\alpha}}^{\;\;{\gamma}}\nabla_{\gamma}
+\frac{\imath}{4}\omega_{{\alpha}}^{\;\;cd}({\sigma}_{dc})_{{\beta}}^{\;\;{\gamma}}\bar{\nabla}_{\gamma}\nonumber
\\ \nonumber
&&+\left[-\frac{1}{2}C_{{\alpha}\bar{{\beta}}}^{\;\;\;\;\;C}\omega_{C}^{\;\;\;\;cd}+\frac{1}{2}E_{{\alpha}}\bar{\omega}_{{{\beta}}}^{\;\;\;cd}
+\frac{1}{2}\bar{E}_{{\beta}}\omega_{{{\alpha}}}^{\;\;\;cd}+\frac{1}{2}
\omega_{{{\alpha}}\;\;}^{\;\;b[c}\;\bar{\omega}_{{{\beta}}b}^{\;\;\;\;d]}\right]M_{dc}
\\ \nonumber
&&+\left[-C_{{\alpha}\bar{{\beta}}}^{\;\;\;\;\;C}{\Gamma}_C+E_{{\alpha}}\bar{{\Gamma}}_{{\beta}}+\bar{E}_{{\beta}}{\Gamma}_{{\alpha}}\right]Z,\\
\end{eqnarray}
\begin{eqnarray}
\label{ALG4}
[\nabla_{\alpha},{\nabla}_b]&=&C_{{\alpha}{b}}^{\;\;\;\;C}\nabla_C-{\omega}_{{\alpha}{b}}^{\;\;\;\;c}\nabla_c
-\frac{\imath}{4}\omega_{b}^{\;\;cd}({\sigma}_{dc})_{{\alpha}}^{\;\;{\gamma}}{\nabla}_{\gamma}\nonumber
\\ \nonumber
&&+\frac{1}{2}\left[-C_{{\alpha}{b}}^{\;\;\;\;\;C}\omega_{C}^{\;\;\;\;cd}+E_{{\alpha}}{\omega}_{{b}}^{\;\;\;cd}
-{E}_{b}\omega_{{{\alpha}}}^{\;\;\;cd}-
\omega_{{{\alpha}{e}}\;\;}^{\;\;\;\;[c}\;{\omega}_{{b}}^{\;\;d]e}\right]M_{dc}
\\ \nonumber
&&+\left[-C_{{\alpha}{b}}^{\;\;\;\;C}{\Gamma}_C+E_{{\alpha}}{{\Gamma}}_{b}-{E}_{b}{\Gamma}_{{\alpha}}\right]Z.\\
\end{eqnarray}
Compering the algebra (\ref{ALG1})-(\ref{ALG4}) from (\ref{lub5}),
the super torsion components can be identified in terms of the
super anholonomy and super spin conection components, as it is
shown below
\begin{eqnarray}
\label{lubf26}
T_{a{b}}^{\;\;\;\;c}=C_{a{b}}^{\;\;\;\;c}+\omega_{b{a}}^{\;\;\;\;c}-\omega_{a{b}}^{\;\;\;\;c};\;\;\;\;T_{a{b}}^{\;\;\;\;{\gamma}}=C_{a{b}}^{\;\;\;\;{\gamma}};\;\;\;\;
T_{a{b}}^{\;\;\;\;\bar{{\gamma}}}=C_{a{b}}^{\;\;\;\;\bar{{\gamma}}}.\nonumber
\end{eqnarray}
\begin{eqnarray}
\label{lubf17} T_{{\alpha}{\beta}}^{\;\;\;c}=C_{{\alpha}{\beta}}^{\;\;\;\;c};\;\;\;
T_{{\alpha}{\beta}}^{\;\;\;\;{\gamma}}=C_{{\alpha}{\beta}}^{\;\;\;\;{\gamma}}+\frac{\imath}{4}\omega_{{{\beta}}cd}({\sigma}^{dc})_{\alpha}^{\;\;{\gamma}}+
\frac{\imath}{4}\omega_{{{\alpha}}cd}({\sigma}^{dc})_{\beta}^{\;\;{\gamma}};\;\;\;
T_{{\alpha}{\beta}}^{\;\;\;\;\bar{{\gamma}}}=C_{{\alpha}{\beta}}^{\;\;\;\;\bar{{\gamma}}}.\nonumber
\end{eqnarray}
\begin{eqnarray}
\label{lubf20}
T_{{\alpha}\bar{{\beta}}}^{\;\;\;\;c}=C_{{\alpha}\bar{{\beta}}}^{\;\;\;\;c};\;\;\;
T_{{\alpha}\bar{{\beta}}}^{\;\;\;\;{\gamma}}=C_{{\alpha}\bar{{\beta}}}^{\;\;\;\;{\gamma}}+\frac{\imath}{4}\omega_{\bar{{\beta}}cd}({\sigma}^{dc})_{\alpha}^{\;\;{\gamma}};\;\;\;
T_{{\alpha}\bar{{\beta}}}^{\;\;\;\;\bar{{\gamma}}}=C_{{\alpha}\bar{{\beta}}}^{\;\;\;\;\bar{{\gamma}}}+\frac{\imath}{4}\omega_{{{\alpha}}cd}({\sigma}^{dc})_{\beta}^{\;\;{\gamma}}.\nonumber
\end{eqnarray}
\begin{eqnarray}
\label{torsions}
T_{{\alpha}{b}}^{\;\;\;\;c}=C_{{\alpha}{b}}^{\;\;\;\;c}-\omega_{{\alpha}{b}}^{\;\;\;\;c};\;\;\;\;\;\;
T_{{\alpha}{b}}^{\;\;\;\;{\gamma}}=C_{{\alpha}{b}}^{\;\;\;\;{\gamma}}+\frac{\imath}{4}\omega_{bcd}({\sigma}^{cd})_{\alpha}^{\;\;{\gamma}};\;\;\;\;\;
T_{{\alpha}{b}}^{\;\;\;\;\bar{{\gamma}}}=C_{{\alpha}{b}}^{\;\;\;\;\bar{{\gamma}}}.
\end{eqnarray}

In order to eliminate the spin connections as independent fields,
it is necessary to impose some restrictions on the torsion
superfield. To accomplish this, the following suitable constraints
are considered, through which we are able to write the spin
connection in terms of the anholonomy:
\begin{equation}\label{const1}
T_{ab}^{\;\;\;c}=0\;\Rightarrow\;\omega_{abc}=\frac{1}{2}\left(C_{abc}-C_{acb}-C_{bca}\right);
\end{equation}
\begin{equation}
\label{const2}
T_{{\alpha}{b}}^{\;\;\;\;c}=0\;\Rightarrow\;\omega_{{\alpha}{b}c}=C_{{\alpha}{b}c},
\end{equation}
leaving thus the spin connections as dependent fields. It is worth
noticing that keeping in mind general relativity (torsion free
theory) as a low energy limit of $SUGRA$, the constraint
(\ref{const1}) seems a ``natural" choice. On the other hand, to keep the flat supergeometry ($SUSY$), represented by the algebra shown in (\ref{lub2}), as a particular solution of this curve supergeometry ($SUGRA$), it is necessary that the superspace satisfies the following restriction
\begin{eqnarray}
\label{susy}
T_{{\alpha}\bar{{\beta}}}^{\;\;\;\;c}=({\gamma}^c)_{{\alpha}{\beta}}
\end{eqnarray}
Finally to ensure the existence of (anti)chiral scalar superfields in supergravity, we have to impose a generalization of $D_\alpha\,\bar{\chi}=0$ in curved superspace. This is accomplished through $\nabla_\alpha\bar{\chi}=0$, which means
\begin{equation}
\{\nabla_\alpha,\nabla_\beta\}\bar{\chi}=0.
\end{equation}
Hence
\begin{equation}
\{\nabla_\alpha,\nabla_\beta\}\bar{\chi}=T_{\alpha\beta}^{\;\;\;\;\;C}\nabla_C\bar{\chi}+\frac{1}{2}R_{\alpha\beta{c}}^{\;\;\;\;\;\;\;d}M_d^{\;\;\;c}\bar{\chi}+F_{\alpha\beta}Z\bar{\chi}=0.
\end{equation}
Therefore we have an additional set of constraints, the so called representation preserving constraints, given by
\begin{eqnarray}
\label{rpconstraints}
T_{\alpha\beta}^{\;\;\;\;\;c}=0;\;\;\;T_{\alpha\beta}^{\;\;\;\;\;\gamma}=0;\;\;\;T_{\alpha\beta}^{\;\;\;\;\;\bar{\gamma}}=0.
\end{eqnarray}
The constraints shown in (\ref{const1})-(\ref{susy}) are the so called conventional constraints, 
which essentially allow us to eliminate the spin superconnection as independent field and to keep $SUSY$ 
as a particular solution. These two set of constraint, together with the representation preserving constraint shown in (\ref{rpconstraints}), are called the conformal constraints of the theory, whose corresponding supergeometry is the conformal supergravity.

\section{Perturbation around the flat superspace}
$~~~$
The supergravity theory we are building up is represented by the
algebra (\ref{ALG1})-(\ref{ALG4}), which explicitly gives the
field strengths and curvature of the theory. After the spin
connection is eliminated as independent field, this algebra
essentially depends of the anholonomy. Then the following logical
step is to find an specific form for all components of the
anholonomy superfield in terms of simpler functions.
When this is accomplished, the constructed $SUGRA$ theory will be
described by these functions, which will contain all
the basic physical information. To carry out this, we need to
provide an specific structure to the super vielbein $E_A^{\;\;M}$
using all ``fundamental" geometric objects which appear in
$D=5\;{\cal N}=1$ $SUSY$, namely, those given by (\ref{fundamental}).

First of all let us consider
\begin{equation}
\label{lub41}E_A=E_A^{\;\;M}D_M,
\end{equation}
and let us start considering its vectorial component, which is
writing as
\begin{equation}
\label{lub42}E_a=E_a^{\;\;M}D_M=E_a^{\;\;m}{\partial}_m+E_a^{\;\;\mu}D_\mu+\bar{E}_a^{\;\;\mu}\bar{D}_\mu,
\end{equation}
now expanding $E_a^{\;\;m}$ around the flat solution
${\delta}_a^{\;\;m}$ we have
\begin{equation}
\label{lub43}E_a^{\;\;m}={\delta}_a^{\;\;m}+H_a^{\;\;m},
\end{equation}
hence finally we obtain the perturbative version of (\ref{lub42})
\begin{equation}
\label{lub44}E_a={\partial}_a+H_a^{\;\;m}{\partial}_m+H_a^{\;\;\mu}D_\mu+\bar{H}_a^{\;\;\mu}\bar{D}_\mu.
\end{equation}
It is not complicated to realize that the fields $H_a^{\;\;m}$,
$H_a^{\;\;\mu}$ and its conjugated $\bar{H}_a^{\;\;\mu}$ cannot be
expressed in terms of the fundamental geometric objects given by
(\ref{fundamental}), thus in some sense they are considered as
fundamental objects of the theory. Indeed these field are
identified as the graviton $H_a^{\;\;m}$ and the gravitino
$H_a^{\;\;\mu}$ with its ``conjugated" $\bar{H}_a^{\;\;\mu}$.

On the other hand, the spinorial component $E_{\alpha}$ of the
superfield $E_A$ is written as
\begin{equation}
\label{lub45}E_{\alpha}=E_{\alpha}^{\;\;M}D_M=E_{\alpha}^{\;\;m}{\partial}_m+E_{\alpha}^{\;\;\mu}D_\mu+\bar{E}_{\alpha}^{\;\;\mu}\bar{D}_\mu,
\end{equation}
again expanding $E_{\alpha}^{\;\;\mu}$ around the flat solution
${\delta}_{\alpha}^{\;\;\mu}$ we have
\begin{equation}
\label{lub46}E_{\alpha}^{\;\;\mu}={\delta}_{\alpha}^{\;\;\mu}+H_{\alpha}^{\;\;\mu},
\end{equation}
hence we obtain the perturbative version of (\ref{lub45})
\begin{equation}
\label{lub47}E_{\alpha}=D_{\alpha}+H_{\alpha}^{\;\;\mu}D_\mu+\bar{H}_{\alpha}^{\;\;\mu}\bar{D}_\mu+H_{\alpha}^{\;\;m}{\partial}_m.
\end{equation}
The fields $H_{\alpha}^{\;\;\mu}$ and $\bar{H}_{\alpha}^{\;\;\mu}$ can be
expressed as a linear combination of the fundamental objects
(\ref{fundamental}) by
\begin{eqnarray}
\label{lub51}
H_{\alpha}^{\;\;\mu}={\delta}_{\alpha}^{\;\;\mu}\psi^1+\imath({\gamma}^a)_{\alpha}^{\;\;\mu}\psi_a^1+\frac{1}{4}({\sigma}^{ab})_{\alpha}^{\;\;\mu}\psi_{ab}^1
\end{eqnarray}
and
\begin{eqnarray}
\label{lub52}
\bar{H}_{\alpha}^{\;\;\mu}={\delta}_{\alpha}^{\;\;\mu}\psi^2+\imath({\gamma}^a)_{\alpha}^{\;\;\mu}\psi_a^2+\frac{1}{4}({\sigma}^{ab})_{\alpha}^{\;\;\mu}\psi_{ab}^2.
\end{eqnarray}
The coefficients $\psi$'s are the so called semi-prepotentials of the
theory. We will see later that it is possible to obtain an
explicit form to some semi-prepotentials in terms of $H$'s
fields by using the constraints on supertorsion
components.

Using (\ref{lub51}) and (\ref{lub52}) in (\ref{lub47}) we obtain
the spinorial components of the superspace derivative in terms of
the semi-prepotentials $\psi$'s and the fields
$H_{\alpha}^{\;\;m}$
\begin{eqnarray}
\label{lub53}
E_{\alpha}&=D_{\alpha}+\left[{\delta}_{\alpha}^{\;\;\mu}\psi^1+\imath({\gamma}^a)_{\alpha}^{\;\;\mu}\psi_a^1+\frac{1}{4}({\sigma}^{ab})_{\alpha}^{\;\;\mu}\psi_{ab}^1\right]D_\mu
\nonumber \\
&+\left[{\delta}_{\alpha}^{\;\;\mu}\psi^2+\imath({\gamma}^a)_{\alpha}^{\;\;\mu}\psi_a^2+\frac{1}{4}({\sigma}^{ab})_{\alpha}^{\;\;\mu}\psi_{ab}^2\right]\bar{D}_\mu+H_{\alpha}^{\;\;m}
\partial_m,
\end{eqnarray}
hence
\begin{eqnarray}
\label{lub54}
\bar{E}_{\alpha}&=\bar{D}_{\alpha}+\left[{\delta}_{\alpha}^{\;\;\mu}({\psi}^2)^*-\imath({\gamma}^a)_{\alpha}^{\;\;\mu}({\psi}_a^2)^*-\frac{1}{4}({\sigma}^{ab})_{\alpha}^{\;\;\mu}({\psi}_{ab}^2)^*\right]D_\mu
\nonumber \\
&+\left[{\delta}_{\alpha}^{\;\;\mu}({\psi}^1)^{*}-\imath({\gamma}^a)_{\alpha}^{\;\;\mu}({\psi}_a^1)^{*}-\frac{1}{4}({\sigma}^{ab})_{\alpha}^{\;\;\mu}({{\psi}_{ab}^1})^{*}\right]\bar{D}_\mu+\bar{H}_{\alpha}^{\;\;m}
\partial_m;
\end{eqnarray}

Now using (\ref{lub44}), (\ref{lub53}) and (\ref{lub54}) in
(\ref{anholonomy}) and keeping linear terms, we are able to
express all the anholonomy components in terms of the graviton,
gravitino, the semi-prepotential fields $\psi$'s, and the
fields $H_{\alpha}^{\;\;m}$. Hence we have
\begin{equation}
\label{10}
C_{{a}b}^{\;\;\;\;c}=\partial_aH_b^{\;\;c}-\partial_bH_a^{\;\;c};
\end{equation}
\begin{equation}
\label{11}
C_{{a}b}^{\;\;\;\;{\gamma}}=\partial_aH_b^{\;\;{\gamma}}-\partial_bH_a^{\;\;{\gamma}};
\end{equation}
\begin{equation}
\label{12}
C_{{a}b}^{\;\;\;\;\bar{{\gamma}}}=\partial_a\bar{H}_b^{\;\;{\gamma}}-\partial_b\bar{H}_a^{\;\;{\gamma}}.
\end{equation}

\begin{equation}
\label{1}
C_{{\alpha}{\beta}}^{\;\;\;\;c}=\left[{\delta}_{\alpha}^{\;\;{\gamma}}\psi^2+\imath({\gamma}^a)_{\alpha}^{\;\;{\gamma}}\psi_a^2+\frac{1}{4}({\sigma}^{ab})_{\alpha}^{\;\;{\gamma}}\psi_{ab}^2\right]
({\gamma}^c)_{{\beta}{\gamma}}+D_{{\alpha}}H_{\beta}^{\;\;c}+({\alpha}\leftrightarrow{\beta});
\end{equation}
\begin{equation}
\label{2}
C_{{\alpha}{\beta}}^{\;\;\;\;{\gamma}}=D_{\alpha}\left[{\delta}_{\beta}^{\;\;{\gamma}}\psi^1+\imath({\gamma}^a)_{\beta}^{\;\;{\gamma}}\psi_a^1+\frac{1}{4}({\sigma}^{ab})_{\beta}^{\;\;{\gamma}}\psi_{ab}^1\right]
+({\alpha}\leftrightarrow{\beta});
\end{equation}
\begin{equation}
\label{3}
C_{{\alpha}{\beta}}^{\;\;\;\;\bar{{\gamma}}}=D_{\alpha}\left[{\delta}_{\beta}^{\;\;{\gamma}}\psi^2+\imath({\gamma}^a)_{\beta}^{\;\;{\gamma}}\psi_a^2+\frac{1}{4}({\sigma}^{ab})_{\beta}^{\;\;{\gamma}}\psi_{ab}^2\right]
+({\alpha}\leftrightarrow{\beta});
\end{equation}
\begin{eqnarray}
\label{4}
C_{{\alpha}\bar{{\beta}}}^{\;\;\;\;c}=\eta_{{\alpha}{\beta}}\left[\imath\eta^{mc}((\psi_m^1)^*-\psi_m^1)+\frac{1}{4}\eta^{\mu\nu}(D_\mu\bar{H}_\nu^{\;\;c}+\bar{D}_{\nu}H_\mu^{\;\;c})\right]
\nonumber \\
+({\gamma}^a)_{{\alpha}{\beta}}\left[{\delta}_a^{\;\;c}(1+\psi^1+(\psi^1)^*)-H_a^{\;\;c}+\frac{\imath}{4}\eta^{c[m}{\delta}_a^{\;n]}(\psi_{mn}^1-(\psi_{mn}^1)^*)+X_a^{\;\;c}
\right]
\nonumber \\
+({\sigma}_{ab})_{{\alpha}{\beta}}\left[\frac{1}{2}\eta^{m[a}\eta^{b]c}(\psi_m^1+(\psi_m^1)^*)-\frac{1}{8}\epsilon^{mncab}(\psi_{mn}^1+(\psi_{mn}^1)^*)+X^{cab}
\right];
\end{eqnarray}
\begin{equation}
\label{Xmc} X_a^{\;\;c}\equiv
-\frac{1}{4}({\gamma}_a)^{{\alpha}{\beta}}(D_{\alpha}\bar{H}_{\beta}^{\;\;c}+\bar{D}_{{\beta}}H_{\alpha}^{\;\;c});
\end{equation}
\begin{equation}
\label{Xmc} X^{cab}\equiv
-\frac{1}{8}({\sigma}^{ab})^{{\alpha}{\beta}}(D_{\alpha}\bar{H}_{\beta}^{\;\;c}+\bar{D}_{{\beta}}H_{\alpha}^{\;\;c});
\end{equation}
\begin{eqnarray}
\label{5} C_{{\alpha}\bar{{\beta}}}^{\;\;\;\;{{\gamma}}}=
D_{\alpha}\left[{\delta}_{\beta}^{\;\;{\gamma}}({\psi}^2)^*-\imath({\gamma}^a)_{\beta}^{\;\;{\gamma}}({\psi}_a^2)^*-\frac{1}{4}({\sigma}^{ab})_{\beta}^{\;\;{\gamma}}({\psi}_{ab}^2)^*\right]
\nonumber \\
+\bar{D}_{\beta}\left[{\delta}_{\alpha}^{\;\;{\gamma}}\psi^1+\imath({\gamma}^a)_{\alpha}^{\;\;{\gamma}}\psi_a^1
+\frac{1}{4}({\sigma}^{ab})_{\alpha}^{\;\;{\gamma}}\psi_{ab}^1\right]-({\gamma}^a)_{{\alpha}{\beta}}H_a^{\;\;{\gamma}};&
\end{eqnarray}
\begin{eqnarray}
\label{6}
C_{{\alpha}\bar{{\beta}}}^{\;\;\;\;\bar{{\gamma}}}=D_{\alpha}\left[{\delta}_{\beta}^{\;\;{\gamma}}({\psi}^1)^{*}
-\imath({\gamma}^a)_{\beta}^{\;\;{\gamma}}({\psi}_a^1)^{*}-\frac{1}{4}({\sigma}^{ab})_{\beta}^{\;\;{\gamma}}({\psi}_{ab}^1)^*\right]
\nonumber \\
+\bar{D}_{\beta}\left[{\delta}_{\alpha}^{\;\;{\gamma}}\psi^2+\imath({\gamma}^a)_{\alpha}^{\;\;{\gamma}}\psi_a^2
+\frac{1}{4}({\sigma}^{ab})_{\alpha}^{\;\;{\gamma}}\psi_{ab}^2\right]-({\gamma}^a)_{{\alpha}{\beta}}\bar{H}_a^{\;\;{\gamma}};
\end{eqnarray}

\begin{equation}
\label{7}
C_{{{\alpha}}b}^{\;\;\;\;c}=\bar{H}_b^{\;\;{\gamma}}({\gamma}^c)_{{\alpha}{\gamma}}+D_{{\alpha}}H_b^{\;\;c}-\partial_bH_{\alpha}^{\;\;c};
\end{equation}
\begin{equation}
\label{8}
C_{{{\alpha}}b}^{\;\;\;\;{\gamma}}=D_{\alpha}{H}_b^{\;\;{\gamma}}-\partial_b\left[{\delta}_{\alpha}^{\;\;{\gamma}}\psi^1+\imath({\gamma}^a)_{\alpha}^{\;\;{\gamma}}\psi_a^1+\frac{1}{4}({\sigma}^{ac})_{\alpha}^{\;\;{\gamma}}\psi_{ac}^1\right];
\end{equation}
\begin{equation}
\label{9}
C_{{{\alpha}}b}^{\;\;\;\;\bar{{\gamma}}}=D_{\alpha}\bar{H}_b^{\;\;{\gamma}}-\partial_b\left[{\delta}_{\alpha}^{\;\;{\gamma}}\psi^2+\imath({\gamma}^a)_{\alpha}^{\;\;{\gamma}}\psi_a^2+\frac{1}{4}({\sigma}^{ac})_{\alpha}^{\;\;{\gamma}}\psi_{ac}^2\right];
\end{equation}

With all the components of the anholonomy written in terms of the
semi-prepotential fields $\psi$'s and fields $H$'s, the
next step will be to use some suitable constraint to write the
$\psi$'s fields in terms of $H$'s fields. We will see that a direct consequence of keeping $SUSY$ as a particular solution allow us to determinate the semi-prepotentials $\psi^1_a$ and $\psi_{ab}^1$ in terms of $H_\alpha^{\;\;b}$ and its conjugate, and that the existence of (anti)chiral scalar superfield allow us to determinate the semi-prepotentials $\psi^1_a$ and $\psi_{ab}^1$ in terms of $H_\alpha^{\;\;b}$. Let us start considering $T_{{\alpha}\bar{{\beta}}}^{\;\;\;\;c}=({\gamma}^c)_{{\alpha}{\beta}}$, a "rigid constraint" given in
(\ref{susy}),
which by (\ref{torsions}) means
\begin{equation}
\label{const33} C_{{\alpha}\bar{{\beta}}}^{\;\;\;\;c}=({\gamma}^c)_{{\alpha}{\beta}}.
\end{equation}
Using (\ref{4}) in the expression (\ref{const33}), we obtain
\begin{eqnarray}
\label{rigidconst}
({\gamma}^c)_{{\alpha}{\beta}}=\eta_{{\alpha}{\beta}}\left[\imath\eta^{mc}((\psi_m^1)^*-\psi_m^1)+\frac{1}{4}\eta^{\mu\nu}(D_\mu\bar{H}_\nu^{\;\;c}+\bar{D}_{\nu}H_\mu^{\;\;c})\right]
\nonumber \\
+({\gamma}^a)_{{\alpha}{\beta}}\left[{\delta}_a^{\;\;c}(1+\psi^1+(\psi^1)^*)-H_a^{\;\;c}+\frac{\imath}{4}\eta^{c[m}\delta_a^{\;\;n]}(\psi_{mn}^1-(\psi_{mn}^1)^*)+X_a^{\;\;c}
\right]
\nonumber \\
+({\sigma}_{ab})_{{\alpha}{\beta}}\left[\frac{1}{2}\eta^{m[a}\eta^{b]c}(\psi_m^1+(\psi_m^1)^*)-\frac{1}{8}\epsilon^{mncab}(\psi_{mn}^1+(\psi_{mn}^1)^*)+X^{cab}
\right],
\end{eqnarray}
showing thus that the rigid constraint leads to the
following three independent equations
\begin{equation}
\label{rigidconst1}
\imath\eta^{mc}((\psi_m^1)^*-\psi_m^1)+\frac{1}{4}\eta^{\mu\nu}(D_\mu\bar{H}_\nu^{\;\;c}+\bar{D}_{\nu}H_\mu^{\;\;c})=0,
\end{equation}
\begin{equation}
\label{rigidconst2}
{\delta}_a^{\;\;c}(\psi^1+(\psi^1)^*)-H_a^{\;\;c}+\frac{\imath}{4}\eta^{c[m}\delta_a^{\;\;n]}(\psi_{mn}^1-(\psi_{mn}^1)^*)+X_a^{\;\;c}
=0,
\end{equation}
\begin{equation}
\label{rigidconst3}
\frac{1}{2}\eta^{m[a}\eta^{b]c}(\psi_m^1+(\psi_m^1)^*)-\frac{1}{8}\epsilon^{mncab}(\psi_{mn}^1+(\psi_{mn}^1)^*)+X^{cab}=0.
\end{equation}
From the equations (\ref{rigidconst1}) we have
\begin{equation}
\label{psidsd}
\psi_a^1-(\psi_a^1)^{*}=-\imath\frac{1}{4}\eta_{ac}\eta^{{\alpha}{\beta}}(D_{\alpha}\bar{H}_{\beta}^{\;\;c}+\bar{D}_{{\beta}}H_{\alpha}^{\;\;c}),
\end{equation}
and from the equation (\ref{rigidconst2}) we obtain 
\begin{equation}
\psi^1+(\psi^1)^{*}=\frac{1}{5}[H_a^{\;a}+\frac{1}{4}({\gamma}_a)^{{\alpha}{\beta}}(D_{\alpha}\bar{H}_{\beta}^{\;\;a}+\bar{D}_{{\beta}}H_{\alpha}^{\;\;a})]
\end{equation}
and
\begin{equation}
\label{psiab1conj2}
\psi_{ab}^1-(\psi_{ab}^1)^{*}=-{\imath}\eta_{c[a}H_{b]}^{\;\;c}-\imath\frac{1}{4}\eta_{c[a}({\gamma}_{b]})^{{\alpha}{\beta}}(D_{\alpha}\bar{H}_{\beta}^{\;\;c}
+\bar{D}_{{\beta}}H_{\alpha}^{\;\;c}).
\end{equation}
From (\ref{rigidconst3}) it is found the following two expressions
\begin{equation}
\label{psia1-22}
\psi_a^1+(\psi_a^1)^{*}=\frac{1}{16}({\sigma}_{ac})^{{\alpha}{\beta}}(D_{\alpha}\bar{H}_{\beta}^{\;\;c}+\bar{D}_{{\beta}}H_{\alpha}^{\;\;c});
\end{equation}
\begin{equation}
\label{psiab1}
\psi_{ab}^1+(\psi_{ab}^1)^{*}=\frac{1}{12}\epsilon_{abckl}({\sigma}^{kl})^{{\alpha}{\beta}}(D_{\alpha}\bar{H}_{\beta}^{\;\;c}+\bar{D}_{{\beta}}H_{\alpha}^{\;\;c}).
\end{equation}
Thus from (\ref{psidsd}) and (\ref{psia1-22}) we obtain
\begin{equation}
\label{psia1xx}
\psi_a^1=\frac{1}{8}\left[-\imath\eta_{ac}\eta^{{\alpha}{\beta}}+\frac{1}{4}({\sigma}_{ac})^{{\alpha}{\beta}}\right]
(D_{\alpha}\bar{H}_{\beta}^{\;\;c}+\bar{D}_{{\beta}}H_{\alpha}^{\;\;c}),
\end{equation}
and from (\ref{psiab1conj2}) and (\ref{psiab1}) we have
\begin{eqnarray}
\label{psiab1fin}
\psi_{ab}^1=-{\imath}\frac{1}{2}\eta_{c[a}H_{b]}^{\;\;c}+\frac{1}{8}\left[-\imath\eta_{c[a}({\gamma}_{b]})^{{\alpha}{\beta}}+\frac{1}{3}\epsilon_{abckl}({\sigma}^{kl})^{{\alpha}{\beta}}\right]
(D_{\alpha}\bar{H}_{\beta}^{\;\;c}+\bar{D}_{{\beta}}H_{\alpha}^{\;\;c}), \nonumber \\
\end{eqnarray}
thus the graviton may be expressed in terms of the semi-prepotential $\psi_{ab}^1$ as
\begin{eqnarray}
\label{graviton}
H_{a}^{\;\;d}=-{\imath}\eta^{bd}\psi_{ab}^1+{\imath}\frac{1}{8}\eta^{bd}\left[-\imath\eta_{c[a}({\gamma}_{b]})^{{\alpha}{\beta}}+\frac{1}{3}\epsilon_{abckl}({\sigma}^{kl})^{{\alpha}{\beta}}\right]
(D_{\alpha}\bar{H}_{\beta}^{\;\;c}+\bar{D}_{{\beta}}H_{\alpha}^{\;\;c}). \nonumber \\
\end{eqnarray}

In order to obtain $\psi_a^2$ and $\psi_{ab}^2$, the "chiral constraint" $T_{{\alpha}{{\beta}}}^{\;\;\;\;c}=0$ shown in (\ref{rpconstraints}) is used, which leads to
\begin{equation}
\label{constcg} C_{{\alpha}{{\beta}}}^{\;\;\;\;c}=0.
\end{equation}
Thus using the expression (\ref{1}) in the condition
(\ref{constcg}), we finally obtain
\begin{equation}
\label{psia2}
\psi_a^2=\frac{1}{16}({\sigma}_{ca})^{{\alpha}{\beta}}D_{{\alpha}}H_{\beta}^{\;\;c};
\end{equation}
\begin{equation}
\label{psiab2}
\psi_{ab}^2=-\frac{1}{12}\epsilon_{abcde}({\sigma}^{de})^{{\alpha}{\beta}}D_{{\alpha}}H_{{\beta}}^{\;\;c}.
\end{equation}
The remaining two constraints associate to the chiral representation, that is, $T_{\alpha{{\beta}}}^{\;\;\;\;\;\gamma}=0$
and $T_{\alpha{{\beta}}}^{\;\;\;\;\;\bar{\gamma}}=0,$ lead respectively to
\begin{equation}
\label{chiral5}
C_{{\alpha}{\beta}}^{\;\;\;\;{\gamma}}=
\frac{\imath}{4}\omega_{{{\alpha}}cd}({\sigma}^{cd})_{\beta}^{\;\;{\gamma}}+\frac{\imath}{4}\omega_{{{\beta}}cd}({\sigma}^{cd})_{\alpha}^{\;\;{\gamma}};\;\;\;
\end{equation}
\begin{equation}
\label{chiral6}
C_{{\alpha}{\beta}}^{\;\;\;\;\bar{{\gamma}}}=0,
\end{equation}
where the spin connection component $\omega_{{{\alpha}}bc}$ is given by
\begin{equation}
\label{spin}
\omega_{{{\alpha}}b}^{\;\;\;\;c}=\bar{H}_{b}^{\;\;\gamma}(\gamma^c)_{\alpha\gamma}+D_{\alpha}H_b^{\;\;c}-\partial_{b}H_{\alpha}^{\;\;c}.
\end{equation}
Using (\ref{spin}) in (\ref{chiral5})  we have
\begin{eqnarray}
\label{chiral7}
20D_\alpha\psi^1&=&\imath(\sigma_{ab}\gamma^c\sigma^{ab})^\beta_{\;\;\alpha}D_\beta\psi_c^1+\frac{1}{4}(\sigma_{ab}\sigma^{cd}\sigma^{ab})^\beta_{\;\;\alpha}D_\beta\psi_{cd}^1
\nonumber \\
&&+\imath\frac{1}{4}(\sigma_{ab}\sigma_d^{\;\;\;c}\sigma^{ab})^\beta_{\;\;\alpha}\left[\bar{H}_c^{\;\;\gamma}(\gamma^d)_{\alpha\gamma}+D_{\alpha}H_c^{\;\;d}
-\partial_cH_\alpha^{\;\;d}\right],
\end{eqnarray}
and from (\ref{3}) in (\ref{chiral6}) we obtain
\begin{eqnarray}
\label{chiral8}
20D_\alpha\psi^2=\imath(\sigma_{ab}\gamma^c\sigma^{ab})^\beta_{\;\;\alpha}D_\beta\psi_c^2+\frac{1}{4}(\sigma_{ab}\sigma^{cd}\sigma^{ab})^\beta_{\;\;\alpha}D_\beta\psi_{cd}^2.
\end{eqnarray}

\section{The Bianchi identities}

So far we have impose some restrictions on superspace through some constraints on the supertorsion components. It was necessary to impose the constraints (\ref{const1}) and (\ref{const2}) to leave the spin connections as dependent fields of the anholonomy, the constraint (\ref{susy}) to keep rigid supersymmetry as a particular solution, and (\ref{rpconstraints}) to ensure the existence of (anti)chiral scalar superfields in supergravity. When all these constraint are imposed, the geometry of the superspace is restricted, in consequence the Bianchi identities, which can be written by
\begin{equation}
[\nabla_A,[\nabla_B,\nabla_C\}\}+(-1)^{A(B+C)}[\nabla_B,[\nabla_C,\nabla_A\}\}+(-1)^{C(A+B)}[\nabla_C,[\nabla_A,\nabla_B\}\}=0,
\end{equation}
now contain non trivial information. This information can be read by the following three equations
\begin{eqnarray}
\label{bianchi1}
&&\nabla_A\;T_{BC}^{\;\;\;\;\;F}+(-1)^{A(B+C+D)}T_{BC}^{\;\;\;\;\;D}T_{AD}^{\;\;\;\;\;F}+(-1)^{A(B+C)}\frac{1}{2}R_{BC}^{\;\;\;\;\;\;cd}\Phi_{dcA}^{\;\;\;\;\;\;F}
\nonumber \\
&&+(-1)^{A(B+C)}\nabla_B\;T_{CA}^{\;\;\;\;\;F}+(-1)^{C(B+A)+BD}T_{CA}^{\;\;\;\;\;D}T_{BD}^{\;\;\;\;\;F}+(-1)^{C(A+B)}\frac{1}{2}R_{CA}^{\;\;\;\;\;\;cd}\Phi_{dcB}^{\;\;\;\;\;\;F}
\nonumber \\
&&+(-1)^{C(A+B)}\nabla_C\;T_{AB}^{\;\;\;\;\;F}+(-1)^{CD}T_{AB}^{\;\;\;\;\;D}T_{CD}^{\;\;\;\;\;F}+\frac{1}{2}R_{AB}^{\;\;\;\;\;\;cd}\Phi_{dcC}^{\;\;\;\;\;\;F}=0;
\nonumber \\
\end{eqnarray}
\begin{eqnarray}
\label{bianchi2}
&&(-1)^{A(B+C+D)}T_{BC}^{\;\;\;\;\;D}R_{ADc}^{\;\;\;\;\;\;\;d}+\nabla_AR_{BCc}^{\;\;\;\;\;\;\;d}
\nonumber \\
&&+(-1)^{C(A+B)+BD}T_{CA}^{\;\;\;\;\;D}R_{BDc}^{\;\;\;\;\;\;\;d}+\nabla_BR_{CAc}^{\;\;\;\;\;\;\;d}
\nonumber \\
&&+(-1)^{CD}T_{AB}^{\;\;\;\;\;D}R_{CDc}^{\;\;\;\;\;\;\;d}+\nabla_CR_{ABc}^{\;\;\;\;\;\;\;d}
=0;
\end{eqnarray}
\begin{eqnarray}
\label{bianchi3}
&&(-1)^{A(B+C+D)}T_{BC}^{\;\;\;\;\;D}F_{AD}+\nabla_AF_{BC}
\nonumber \\
&&+(-1)^{C(A+B)+BD}T_{CA}^{\;\;\;\;\;D}F_{BD}+(-1)^{A(B+C)}\nabla_BF_{CA}
\nonumber \\
&&+(-1)^{CD}T_{AB}^{\;\;\;\;\;D}F_{CD}+(-1)^{C(A+B)}\nabla_CF_{AB}
=0;
\end{eqnarray}
where
\begin{eqnarray}
\Phi_{abC}^{\;\;\;\;\;\;D} =
\left( {\begin{array}{cc}
 \Phi_{abc}^{\;\;\;\;\;\;d} & 0  \\
 0 & \Phi_{ab{\gamma}}^{\;\;\;\;\;\;{\delta}} \\
 \end{array} } \right) = \left( {\begin{array}{cc}
 \eta_{c[a}\delta_{b]}^{\;\;d} & 0  \\
 0 & \imath\frac{1}{2}(\sigma_{ab})_{\gamma}^{\;\;\delta} \\
 \end{array} } \right). \nonumber
\end{eqnarray}

It is well known \cite{Dragon} that it is sufficient to analyze the Bianchi identities (\ref{bianchi1}) and (\ref{bianchi3}), since all equations contained in (\ref{bianchi2}) are identically satisfied when (\ref{bianchi1}) and (\ref{bianchi3}) hold. Hence using the constraints (\ref{const1}), (\ref{const2}), (\ref{susy}) and (\ref{rpconstraints}) in the Bianchi identities (\ref{bianchi1}) and (\ref{bianchi3}) we will be able to obtain the curvature and field strength superfield components in terms of the smaller set of superfields of the theory.

\section{Symmetries and semi-prepotentials }
$~~~$
In order to obtain some information on $\psi's$, let's see the
behaviour of them under the scale, $U(1)$ and Lorentz Symmetry,
which are represented respectively by
\begin{eqnarray}
\label{symmeties} E_a\rightarrow E'_a=e^{f_0}E_a; \hspace{.5cm}
E_\alpha\rightarrow E'_\alpha=e^{\frac{1}{2}{f_0}}E_\alpha;
\hspace{.5cm} \bar{E}_\alpha\rightarrow
\bar{E}'_\alpha=e^{\frac{1}{2}{f_0}}\bar{E}_\alpha  \\
E_\alpha\rightarrow E'_\alpha=e^{\imath\frac{1}{2}{f_0}}E_\alpha;
\hspace{.5cm} \bar{E}_\alpha\rightarrow
\bar{E}'_\alpha=e^{-\imath\frac{1}{2}{f_0}}\bar{E}_\alpha \\
E_a\rightarrow E'_a=\Lambda_a^{\;\;b}E_b; E_\alpha\rightarrow
E'_\alpha=e^{\frac{1}{8}\Lambda^{ab}(\sigma_{ab})_\alpha^{\;\;\beta}}E_\beta;
 \bar{E}_\alpha\rightarrow
\bar{E}'_\alpha=e^{\frac{1}{8}\Lambda^{ab}(\sigma_{ab})_\alpha^{\;\;\beta}}\bar{E}_\beta\nonumber \\
\end{eqnarray}

Let us begin considering the scale transformation 
\begin{eqnarray}
\label{trans} E_a\rightarrow E'_a=e^{f_0}E_a,
\nonumber \\
E_\alpha\rightarrow E'_\alpha=e^{\frac{1}{2}{f_0}}E_\alpha,
\nonumber \\
\bar{E}_\alpha\rightarrow
\bar{E}'_\alpha=e^{\frac{1}{2}{f_0}}\bar{E}_\alpha.
\end{eqnarray}
Considering the infinitesimal version of (\ref{trans}) and the
perturbative expression of $(E_a,E_\alpha,\bar{E}_\alpha)$ around
the flat solution, we have
\begin{equation}
\label{trans12x} E_a\rightarrow
E'_a=(1+f_0)({\partial}_a+H_a^{\;\;m}{\partial}_m+H_a^{\;\;\mu}D_\mu+\bar{H}_a^{\;\;\mu}\bar{D}_\mu),
\end{equation}
\begin{equation}
\label{trans22x} E_\alpha\rightarrow
E'_\alpha=(1+\frac{1}{2}f_0)E_\alpha,
\end{equation}
\begin{equation}
\label{trans32x} \bar{E}_\alpha\rightarrow
\bar{E}'_\alpha=(1+\frac{1}{2}f_0)\bar{E}_\alpha.
\end{equation}
The Eq. (\ref{trans12x}) can be written as 
\begin{equation}
\label{trans13} E_a\rightarrow
E'_a={\partial}_a+(f_0{\delta}_a^{\;\;m}+H_a^{\;\;m}){\partial}_m+H_a^{\;\;\mu}D_\mu+\bar{H}_a^{\;\;\mu}\bar{D}_\mu,
\end{equation}
showing that there is a shift on $H_a^{\;\;m}$ due to the scale transformation, as shown bellow
\begin{equation}
\label{trans1f} E_a\rightarrow E'_a=e^{f_0}E_a \Rightarrow
H_a^{\;\;m}\rightarrow H_a^{\;\;m}+f_0{\delta}_a^{\;\;m}.
\end{equation}
On the other hand, the Eqs. (\ref{trans22x}) and (\ref{trans32x}) can be written as
\begin{equation}
\label{trans23} E_\alpha\rightarrow
E'_\alpha=E_\alpha+\frac{1}{2}f_0\delta_\alpha^{\;\;\mu}\,D_\mu
\end{equation}
\begin{equation}
\label{trans33} \bar{E}_\alpha\rightarrow
\bar{E}'_\alpha=\bar{E}_\alpha+\frac{1}{2}f_0\delta_\alpha^{\;\;\mu}\,\bar{D}_\mu.
\end{equation}
Now using the explicit form of $E_\alpha$ and
$\bar{E}_\alpha$ given in (\ref{lub53}) and
(\ref{lub54}), we have
\begin{equation}
\label{trans2ff} E_\alpha\rightarrow
E'_\alpha=e^{\frac{1}{2}{f_0}}E_\alpha \Rightarrow
\psi^1\rightarrow \psi^1+\frac{1}{2}f_0,
\end{equation}
\begin{equation}
\label{trans3ff} \bar{E}_\alpha\rightarrow
\bar{E}'_\alpha=e^{\frac{1}{2}{f_0}}\bar{E}_\alpha \Rightarrow
(\psi^1)^{*}\rightarrow (\psi^1)^{*}+\frac{1}{2}f_0,
\end{equation}
showing thus that the scale transformation prodeces a shift on the real parte of the semi-prepotential $\psi^1$. This can be seen more clearly through 
the following useful decomposition
\begin{eqnarray}
\label{redefinitionx}
\psi^1=\frac{1}{2}(\hat\psi^1+\imath\tilde{\psi^1}),
\end{eqnarray}
where $\hat\psi$ and $\tilde{\psi}$ are real functions. Hence we have
\begin{equation}
\label{trans2ffr} E_\alpha\rightarrow
E'_\alpha=e^{\frac{1}{2}{f_0}}E_\alpha \Rightarrow
\frac{1}{2}(\hat\psi^1+\imath\tilde{\psi^1})\rightarrow
\frac{1}{2}(\hat\psi^1+\imath\tilde{\psi^1})+\frac{1}{2}f_0,
\end{equation}
\begin{equation}
\label{trans3ffr} \bar{E}_\alpha\rightarrow
\bar{E}'_\alpha=e^{\frac{1}{2}{f_0}}\bar{E}_\alpha \Rightarrow
\frac{1}{2}(\hat\psi^1-\imath\tilde{\psi^1})\rightarrow
\frac{1}{2}(\hat\psi^1-\imath\tilde{\psi^1})+\frac{1}{2}f_0,
\end{equation}
hence 
\begin{equation}
\hat\psi^1\rightarrow\hat\psi^1+f_0;\;\;\;\imath\tilde{\psi^1}\rightarrow\imath\tilde{\psi^1},
\end{equation}
showhing thus that under the scale transformation there is a shift on
$\hat\psi^1$, leaving invariant $\tilde\psi^1$ in the expresion (\ref{redefinitionx}).

Let us consider now the $U(1)$ transformation
\begin{eqnarray}
\label{transu1} 
E_\alpha\rightarrow E'_\alpha=e^{\imath\frac{1}{2}{f}}E_\alpha,
\nonumber \\
\bar{E}_\alpha\rightarrow
\bar{E}'_\alpha=e^{-\imath\frac{1}{2}{f}}\bar{E}_\alpha.
\end{eqnarray}
Considering the infinitesimal version of (\ref{transu1}) and the
perturbative expression of $(E_\alpha,\bar{E}_\alpha)$ around
the flat solution, we obtain
\begin{equation}
\label{trans23u1} E_\alpha\rightarrow
E'_\alpha=E_\alpha+\imath\frac{1}{2}f\delta_\alpha^{\;\;\mu}{D}_\mu,
\end{equation}
\begin{equation}
\label{trans33u1} \bar{E}_\alpha\rightarrow
\bar{E}'_\alpha=\bar{E}_\alpha-\imath\frac{1}{2}f\delta_\alpha^{\;\;\mu}\bar{D}_\mu.
\end{equation}
Now using the explicit form of $E_\alpha$ and
$\bar{E}_\alpha$ given in (\ref{lub53}) and
(\ref{lub54}), we have
\begin{equation}
\label{trans2ffu1} E_\alpha\rightarrow
E'_\alpha=e^{\imath\frac{1}{2}{f}}E_\alpha \Rightarrow
\psi^1\rightarrow \psi^1+\imath\frac{1}{2}f,
\end{equation}
\begin{equation}
\label{trans3ffu1} \bar{E}_\alpha\rightarrow
\bar{E}'_\alpha=e^{-\imath\frac{1}{2}{f}}\bar{E}_\alpha
\Rightarrow (\psi^1)^{*}\rightarrow (\psi^1)^{*}-\imath\frac{1}{2}f.
\end{equation}
Again, as in the previous case, there is a shift on the semi-prepotential $\psi^1$. In this case the $U(1)$ symmetry produces a shift on 
the imaginary part of the semi-prepotential $\psi^1$. This can be seen clearly using the decomposition shown in Eq. (\ref{redefinitionx}) as following
\begin{equation}
\label{trans2ffry}  E_\alpha\rightarrow
E'_\alpha=e^{\imath\frac{1}{2}{f}}E_\alpha \Rightarrow
\frac{1}{2}(\hat\psi^1+\imath\tilde{\psi^1})\rightarrow
\frac{1}{2}(\hat\psi^1+\imath\tilde{\psi^1})+\imath\frac{1}{2}f,
\end{equation}
\begin{equation}
\label{trans3ffry} \bar{E}_\alpha\rightarrow
\bar{E}'_\alpha=e^{-\imath\frac{1}{2}{f}}\bar{E}_\alpha \Rightarrow
\frac{1}{2}(\hat\psi^1-\imath\tilde{\psi^1})\rightarrow
\frac{1}{2}(\hat\psi^1-\imath\tilde{\psi^1})-\imath\frac{1}{2}f,
\end{equation}
thus
\begin{equation}
\hat\psi^1\rightarrow\hat\psi^1;\;\;\;\imath\tilde{\psi^1}\rightarrow\imath\tilde{\psi^1}+\imath{f},
\end{equation}
hence the $U(1)$ transformation produces a shift on  $\tilde\psi^1$, leaving invariant $\hat\psi^1$ in the expresion (\ref{redefinitionx}) 
for the semi-prepotential $\psi^1$.

Now let's consider the Lorentz transformation on the vector
component $E_a$
\begin{equation}
E_a\rightarrow E'_a=\Lambda_a^{\;\;b}E_b.
\end{equation}
Now we consider the infinitesimal Lorentz transformation
\begin{equation}
\Lambda_a^{\;\;b}=\delta_a^{\;\;b}+\epsilon_a^{\;\;b};\hspace{2cm}\epsilon_{ab}=-\epsilon_{ba}
\end{equation}
acting on the perturbative expression of $E_a$ around the flat
solution
\begin{equation}
E_a\rightarrow
E'_a=(\delta_a^{\;\;b}+\epsilon_a^{\;\;b})({\partial}_b+H_b^{\;\;m}{\partial}_m+H_b^{\;\;\mu}D_\mu+\bar{H}_b^{\;\;\mu}\bar{D}_\mu)
\end{equation}
\begin{equation}
E_a\rightarrow
E'_a={\partial}_a+(\epsilon_a^{\;\;m}+H_a^{\;\;m}){\partial}_m+H_a^{\;\;\mu}D_\mu+\bar{H}_a^{\;\;\mu}\bar{D}_\mu
\end{equation}
Hence
\begin{equation}
E_a\rightarrow E'_a=\Lambda_a^{\;\;b}E_b\Rightarrow
H_a^{\;\;m}\rightarrow H_a^{\;\;m}+\epsilon_a^{\;\;m},
\end{equation}
showing thus that the Lorentz transformation produces a shift on
the graviton.

Now considering the Lorentz transformation on the spinorial
components
\begin{equation}
\label{lor1} E_\alpha\rightarrow
E'_\alpha=e^{\imath\frac{1}{8}\Lambda^{ab}(\sigma_{ab})_\alpha^{\;\;\beta}}E_\beta,
\end{equation}
\begin{equation}
\label{lor2} \bar{E}_\alpha\rightarrow
\bar{E}'_\alpha=e^{-\imath\frac{1}{8}\Lambda^{ab}(\sigma_{ab})_\alpha^{\;\;\beta}}\bar{E}_\beta.
\end{equation}
Considering the infinitesimal transformation we have
\begin{equation}
\label{lor12} E_\alpha\rightarrow
E'_\alpha=[\delta_\alpha^{\;\;\beta}+\imath\frac{1}{8}\Lambda^{ab}(\sigma_{ab})_\alpha^{\;\;\beta}]E_\alpha,
\end{equation}
\begin{equation}
\label{lor22} \bar{E}_\alpha\rightarrow
\bar{E}'_\alpha=[\delta_\alpha^{\;\;\beta}-\imath\frac{1}{8}\Lambda^{ab}(\sigma_{ab})_\alpha^{\;\;\beta}]\bar{E}_\alpha,
\end{equation}
and using the perturbation around the flat solution, we obtain
\begin{equation}
\label{lor13} E_\alpha\rightarrow E'_\alpha=
E_\alpha+\imath\frac{1}{8}\Lambda^{ab}(\sigma_{ab})_\alpha^{\;\;\mu}D_\mu,
\end{equation}
\begin{equation}
\label{lor23} \bar{E}_\alpha\rightarrow \bar{E}'_\alpha=
\bar{E}_\alpha-\imath\frac{1}{8}\Lambda^{ab}(\sigma_{ab})_\alpha^{\;\;\mu}\bar{D}_\mu.
\end{equation}
Now using the explicit form of $E_\alpha$ and
$\bar{E}_\alpha$ given in (\ref{lub53}) and
(\ref{lub54}), we have
\begin{equation}
\label{lor4f} E_\alpha\rightarrow
E'_\alpha=e^{\imath\frac{1}{8}\Lambda^{ab}(\sigma_{ab})_\alpha^{\;\;\beta}}E_\beta\Rightarrow
\psi_{ab}^1\rightarrow \psi_{ab}^1+\imath\frac{1}{2}\Lambda_{ab},
\end{equation}
\begin{equation}
\label{lor4f} \bar{E}_\alpha\rightarrow
\bar{E}'_\alpha=e^{\imath\frac{1}{8}\Lambda^{ab}(\sigma_{ab})_\alpha^{\;\;\beta}}\bar{E}_\beta\Rightarrow
(\psi_{ab}^1)^{*}\rightarrow (\psi_{ab}^1)^{*}-\imath\frac{1}{2}\Lambda_{ab},
\end{equation}
showing thus that the Lorentz transformation prodeces a shift on the imaginary part of the pre-potential $\psi_{ab}^1$.

\section{Conclusions}

$~~~$
Using the notion of superspace, a geometrical approach to five
dimensional ${\cal N}=1$ supergravity theory was discussed in detail.
There
was not used the conventional re\-presentation for Grassmann
variables, based in spinors obeying a pseudo-Majorana reality
condition. Instead, the unconventional representation for the
Grassmann varia\-bles $(\theta^{\mu},\bar{\theta}^{\mu})$ given in
\cite{JimLubna} for a $SUSY$ $D=5$, ${\cal N}=1$ representation, was
successfully extended for a supergravity theory, dispensing with
the use of a $SU(2)$ index to the spinor coordinates of the
superspace.

The components of the torsion and curvature superfield were found
through the super (anti)-commutator of the superspace supergravity
covariant derivative, finding these superfields as function of
both the anholonomy and the spin connection. Imposing suitable
constraints on superspace through some super-torsion components,
the spin connection was written in terms of the anholonomy,
eliminating thus the spin connection as un independent field.

Taking a perturbation around the flat superspace, the components
of the superspace derivative were found as the sum of the ``rigid"
(SUSY) part and perturbative terms. These perturbative terms,
arising from the super vielbein components, were written in terms
of functions when the vectorial component of the
superspace derivative was considered. These functions
were a two vectorial component superfield and its supersymmetric
partner, namely, the graviton and gravitino at quantum level. On
the other hand, when the spinorial component of the superspace
derivative was considered, it was possible to write the
perturbative terms as a linear combination of fundamental
geometric objects of SUSY ${\cal N}=1$, introducing thus the semi-prepotential
of the theory.

Using the perturbative version (linearized theory) of the
superspace derivative, it was possible to find all components of
the super anholonomy in terms of the simpler set of superfields, namely,
the graviton, the gravitino and semi-prepotentials. Demanding
consistence with rigid SUSY and the existence of (anti)chiral scalar superfields in supergravity, some suitables constraints 
on superspace were imposed, then all semi-prepotentials
were written in terms of the smaller set of  superfields of the theory, leaving
the two scalar semi-prepotentials $\psi^1$ and $\psi^2$ superfields. Using the Bianchi identities, three set of equations written in terms of superfields were found. Two of these set containing enough information to determinate the curvature and field strength superfield components in terms of the smaller set of superfields of the theory.

It was explained in detail the behaviour of the semi-prepotentials
under the action of the scale, $U(1)$ and Lorentz Symmetry. It was
found that the scale transformation produces a shift on the real
part of the scalar semi-prepotential $\psi^1$, leaving invariant the scalar semi-prepotential $\psi^2$. A simmilar behaviour was found when the $U(1)$
transformation was considered, where only the semi-prepotential $\psi^1$ was affected, producing a shift on its imaginary part, 
leaving invariant its real sector. Finally,  when the Lorentz
transformation was considered, it was found that there is a shift on the imaginary part of the semi-prepotential $\psi_{ab}^1$, 
leaving invariant its real part as well as the semi-prepotential $\psi_{ab}^2$. 

As mentioned before, this work represents a first stage towards the identification of universal features of the
work of \cite{kuzenko4,kuzenko5}, that is, the identification of all relevant elements independent of compensator choice, and therefore, 
in principle, independent of any chosen superspace geometry.

\section*{Acknowledgments}

The author thanks to Dr. S. J. Gates, Jr. for valuable discussions and suggestions, and
the Center for String \& Particle Theory at the University of Maryland for the hospitality and
financial support. Also thanks to Gabriele Tartaglino-Mazzucchelli for useful discussions. This work was partially supported by {\it Desarrollo Profesoral de la Universidad
Sim\'on Bol\'{\i}var}.

\section*{Appendix}

The action of the generator $M_a^{\;b}$ on spinors $\Psi_{\alpha}$ and
vectors $X_a$ is given by
\begin{equation}
\label{ap1}
[M_{ab},\Psi_{\alpha}]=\frac{\imath}{2}({\sigma}_{ab})_{\alpha}^{\;\;{\gamma}}\Psi_{\gamma};\;\;\;
[M_{ab},X_c]=\eta_{ca}X_b-\eta_{cb}X_a,
\end{equation}
hence
\begin{equation}
\label{ap1}
[M_{ab},\Phi_{{{\alpha}}cd}]=\frac{\imath}{2}({\sigma}_{ab})_{\alpha}^{\;\;{\gamma}}\Phi_{{{\gamma}}cd}+\eta_{ca}\Phi_{{{\alpha}}bd}-\eta_{cb}\Phi_{{{\alpha}}ad}
+\eta_{da}\Phi_{{{\alpha}}cb}-\eta_{db}\Phi_{{{\alpha}}cb}.
\end{equation}

For instance let us consider
\begin{equation}
\label{tor4}
\{\nabla_{\alpha},\nabla_{\beta}\}=C_{{\alpha}{\beta}}^{\;\;\;\;\;C}E_C+\{E_{\alpha},\Upsilon_{\beta}\}+\{\Upsilon_{\alpha},E_{\beta}\}+\{\Upsilon_{\alpha},\Upsilon_{\beta}\},
\end{equation}
thus computing each anticommutator
\begin{equation}
\label{tor5} \{E_{\alpha},\Upsilon_{\beta}\}
=\frac{1}{2}(E_{{\alpha}}\omega_{{{\beta}}c}^{\;\;\;\;d})M_d^{\;\;c}+\frac{\imath}{4}\omega_{{\beta}}^{\;\;cd}({\sigma}_{dc})_{{\alpha}}^{\;\;{\gamma}}E_{\gamma}+E_{{\alpha}}{\Gamma}_{{\beta}}Z,
\end{equation}
\begin{eqnarray}
\label{tor7} \{\Upsilon_{\alpha},\Upsilon_{\beta}\}=
\frac{1}{4}\{\omega_{{{\alpha}}b}^{\;\;\;\;c}M_c^{\;\;b},\omega_{{{\beta}}d}^{\;\;\;\;e}M_e^{\;\;d}\}
+\frac{\imath}{4}\omega_{{{\alpha}}b}^{\;\;\;\;c}({\sigma}_c^{\;\;b})_{\beta}^{\;\;{\gamma}}{\Gamma}_{{\gamma}}Z
+\frac{\imath}{4}\omega_{{{\beta}}d}^{\;\;\;\;e}({\sigma}_e^{\;\;d})_{\alpha}^{\;\;{\gamma}}{\Gamma}_{{\gamma}}Z.\nonumber \\
\end{eqnarray}
Using (\ref{tor5}) and (\ref{tor7}) in (\ref{tor4}) we have
\begin{eqnarray}
\label{tor8}
\{\nabla_{\alpha},\nabla_{\beta}\}=C_{{\alpha}{\beta}}^{\;\;\;\;\;C}E_C+\frac{1}{2}(E_{{\alpha}}\omega_{{{\beta}}c}^{\;\;\;\;d})M_d^{\;\;c}
+\frac{\imath}{4}\omega_{{\beta}}^{\;\;cd}({\sigma}_{dc})_{{\alpha}}^{\;\;{\gamma}}E_{\gamma}+E_{{\alpha}}{\Gamma}_{{\beta}}Z\nonumber
\\ \nonumber
+\frac{1}{2}(E_{{\beta}}\omega_{{{\alpha}}c}^{\;\;\;\;d})M_d^{\;\;c}
+\frac{\imath}{4}\omega_{{\alpha}}^{\;\;cd}({\sigma}_{dc})_{{\beta}}^{\;\;{\gamma}}E_{\gamma}+E_{{\beta}}{\Gamma}_{{\alpha}}Z
\\ \nonumber
+\frac{1}{4}\{\omega_{{{\alpha}}b}^{\;\;\;\;c}M_c^{\;\;b},\omega_{{{\beta}}d}^{\;\;\;\;e}M_e^{\;\;d}\}
+\frac{\imath}{4}\omega_{{{\alpha}}b}^{\;\;\;\;c}({\sigma}_c^{\;\;b})_{\beta}^{\;\;{\gamma}}{\Gamma}_{{\gamma}}Z
\\
+\frac{\imath}{4}\omega_{{{\beta}}d}^{\;\;\;\;e}({\sigma}_e^{\;\;d})_{\alpha}^{\;\;{\gamma}}{\Gamma}_{{\gamma}}Z,
\end{eqnarray}
with
\begin{eqnarray}
\label{tor10}
\Sigma_{{\alpha}{\beta}}=&&\{\omega_{{{\alpha}}b}^{\;\;\;\;c}M_c^{\;\;b},\omega_{{{\beta}}d}^{\;\;\;\;e}M_e^{\;\;d}\}
\nonumber \\
=&&\omega_{{{\alpha}}}^{\;\;\;ba}\omega_{{{\beta}}}^{\;\;\;dc}[M_{ab},M_{cd}]
+\omega_{{{\alpha}}}^{\;\;\;ba}[M_{ab},\omega_{{{\beta}}cd}]M^{dc}
 +\omega_{{{\beta}}}^{\;\;\;ba}[M_{ab},\omega_{{{\alpha}}cd}]M^{dc}
\nonumber \\
=&&4\omega_{{{\alpha}}\;\;c}^{\;\;\;b}\;\omega_{{{\beta}}bd}M^{dc}
+\frac{\imath}{2}\left[({\sigma}_{ab})_{\alpha}^{\;\;{\gamma}}\omega_{{{\beta}}}^{\;\;\;ba}+({\sigma}_{ab})_{\beta}^{\;\;{\gamma}}\omega_{{{\alpha}}}^{\;\;\;ba}\right]\omega_{{{\gamma}}cd}M^{dc}
\end{eqnarray}
and
\begin{equation}
\label{tor13}
E_C=\nabla_C-\frac{1}{2}\omega_{Cc}^{\;\;\;\;d}M_d^{\;\;c}-\Gamma_CZ
\end{equation}
in (\ref{tor8}), we finally obtain
\begin{eqnarray}
\label{tor18}
\{\nabla_{\alpha},\nabla_{\beta}\}=C_{{\alpha}{\beta}}^{\;\;\;\;\;C}\nabla_C+\frac{\imath}{4}\left[\omega_{{\alpha}}^{\;\;cd}({\sigma}_{dc})_{{\beta}}^{\;\;{\gamma}}
+\omega_{{\beta}}^{\;\;cd}({\sigma}_{dc})_{{\alpha}}^{\;\;{\gamma}}\right]\nabla_{\gamma}\nonumber
\\ \nonumber
+\left[-\frac{1}{2}C_{{\alpha}{\beta}}^{\;\;\;\;\;C}\omega_{Cc}^{\;\;\;\;\;d}+\frac{1}{2}E_{{\alpha}}\omega_{{{\beta}}c}^{\;\;\;\;d}
+\frac{1}{2}E_{{\beta}}\omega_{{{\alpha}}c}^{\;\;\;\;d}+
\omega_{{{\alpha}}\;\;c}^{\;\;\;b}\;\omega_{{{\beta}}b}^{\;\;\;\;d}\right]M_{d}^{\;\;c}
\\ \nonumber
+\left[-C_{{\alpha}{\beta}}^{\;\;\;\;\;C}{\Gamma}_C+E_{{\alpha}}{\Gamma}_{{\beta}}+E_{{\beta}}{\Gamma}_{{\alpha}}\right]Z.\\
\end{eqnarray}

\end{document}